\documentclass[twocolumn]{aastex631}

\shorttitle{TTVs of TOI-2818\lowercase{b}}
\shortauthors{McKee et al.}
\graphicspath{{./}{figures/}}

\begin{document}

\title{A Planet Candidate Orbiting near the Hot Jupiter TOI-2818\,b Inferred through Transit Timing}

\correspondingauthor{Brendan J. McKee}
\email{b.mckee@unsw.edu.au}

\author[0000-0001-8421-2833]{Brendan~J.~McKee}
\affiliation{School of Physics, University of New South Wales, Sydney, NSW 2052, Australia}

 \author[0000-0001-7516-8308]{Benjamin~T.~Montet}
\affiliation{School of Physics, University of New South Wales, Sydney, NSW 2052, Australia}
\affiliation{UNSW Data Science Hub, University of New South Wales, Sydney, NSW 2052, Australia}

\author[0000-0001-7961-3907]{Samuel W.\ Yee}
\altaffiliation{51 Pegasi b Fellow}
\affiliation{Department of Astrophysical Sciences, Princeton University, 4 Ivy Lane, Princeton, NJ 08544, USA}
\affiliation{Center for Astrophysics \textbar \ Harvard \& Smithsonian, 60 Garden St, Cambridge, MA 02138, USA}

\author[0000-0001-8732-6166]{Joel D. Hartman}
\affiliation{Department of Astrophysical Sciences, Princeton University, 4 Ivy Lane, Princeton, NJ 08544, USA}

\author[0000-0002-4265-047X]{Joshua N. Winn}
\affiliation{Department of Astrophysical Sciences, Princeton University, 4 Ivy Lane, Princeton, NJ 08544, USA}

\author[0000-0002-1532-9082]{Jorge H. C. Martins}
\affiliation{Instituto de Astrofísica e Ciências do Espaço, Universidade do Porto, Rua das Estrelas, 4150-762 Porto, Portugal}

\author[0000-0003-4920-738X]{Andr\'{e} M. Silva}
\affiliation{Instituto de Astrof\'{i}sica e Ci\^{e}ncias do Espa\c{c}o, Universidade de Lisboa, Campo Grande, 1749-016 Lisboa, Portugal}

\author[0000-0002-6591-5290]{Alexander L. Wallace}
\affiliation{School of Physics and Astronomy, Monash University, Victoria 3800, Australia}




\begin{abstract}

TOI-2818\,b is a hot Jupiter orbiting a slightly evolved G-type star on a 4.04-day orbit that shows transit timing variations (TTVs) suggestive of a decreasing orbital period. In the most recent year of TESS observations, transits were observed $\sim$8 minutes earlier than expected for a constant period. The implied orbital decay rate is $1.35 \pm 0.25$ s yr$^{-1}$, too fast to be explained by tidal dissipation even considering the evolved nature of the host star. Radial velocity monitoring and astrometric data make the possibility of perturbations from a long-period companion unlikely; further Doppler spectroscopy observations can efficiently confirm or rule out such a companion. Apsidal precession due to the tidal distortion of the planet is also physically implausible. The most plausible explanation for the TTVs appears to be gravitational perturbations from a hitherto undetected planet with mass
$\lesssim$$10\,M_\oplus$ that is in (or near) a mean-motion resonance with the hot Jupiter. Such a planet could be responsible for the observed TTVs while avoiding detection with the available radial velocity and transit data.


\end{abstract}



\section{Introduction} \label{sec:intro}


Hot Jupiters can exhibit transit timing variations (TTVs) arising from tidal interactions, apsidal precession, or companion planets or stars. Tidal interactions allow the star and planet to exchange rotational and orbital angular momentum, which for many hot Jupiters is expected to lead to orbital decay \citep{Patra_2020}. Searches for orbital decay in hot Jupiters have yielded a high-confidence detection of tidal decay, WASP-12\,b \citep{Yee_2020} and a second candidate, Kepler-1658\,b \citep{Vissapragada_2022}. WASP-12\,b is a hot Jupiter on a 1-day orbit decaying by $\sim 30$ ms yr$^{-1}$, while Kepler-1658\,b is on a 4-day orbit decaying by $\sim 130$ ms yr$^{-1}$. When interpreted within a simple theory in which the equilibrium tidal bulge is displaced by a constant lag angle, the implied modified tidal quality factors, the ratio between energy stored in tidal distortions and energy dissipated by friction, are $Q'_*\sim10^5$ for WASP-12\,b and $Q'_*\sim10^4$ for Kepler-1658\,b. Such low $Q'_*$ values are only expected if the star is a subgiant or there are nonlinear effects such as wave breaking \citep{Weinberg_2017, Barker_2020}. Tidal inspiral is expected to occur in many systems, although the size of the effect makes unambiguous confirmation difficult \citep{Patra_2020}. 

Apsidal precession can also produce quasi-sinusoidal TTVs \citep{Patra_2017, Yee_2020}. The time interval between transits will oscillate slightly as the line of apsides of the elliptical orbit rotates relative to the line of sight. Steady acceleration of the star toward or away from us, due to a massive long-period companion can also create the illusion of a changing period due to the R{\o}mer effect, also called light travel time delay, as has been proposed for the WASP-4 system \citep{Bouma_2020}, and observed in triple star systems through eclipse timing variations \citep{Borkovits_2016}. 
In many exoplanet systems TTVs are caused by gravitational interactions with nearby or massive planets. Most hot Jupiter systems do not have detectable nearby companions \citep{Hord_2021}, although there are a few counterexamples, such as WASP-47 \citep{Becker_2015}, WASP-84 \citep{Maciejewski_2023}, WASP-132 \citep{Hord_2022}, Kepler-730 \citep{Canas_2019}, and TOI-2000 \citep{Sha_2023}. For WASP-47 \citep{Becker_2015} and TOI-1130 \citep{Huang_2020, Korth_2023}, detectable TTVs have been induced by the nearby companions in these systems. Analysis of hot Jupiter TTVs in the \textit{Kepler} data set suggests that $\sim12\%$ of hot Jupiter systems may have unseen nearby companions \citep{Wu_2023}. 

TTVs can have small amplitudes and long periods, necessitating long observing baselines to detect them. The \textit{TESS} mission \citep{Ricker_2015} has entered its fifth year of operation and is now revisiting southern hemisphere targets four years after first observing them. For stars outside of the continuous viewing zone, this typically means the available data consist of 1--2 months of coverage with two-year gaps between them. Any slow changes in transit timing might be be invisible across a few months, but detectable over several years.

Such is the case for the hot Jupiter orbiting the slightly evolved G-type star TOI-2818. \citet{Yee_2023} discovered TOI-2818\,b, a hot Jupiter on a 4-day orbit, using \textit{TESS} data from Sectors 7, 8, and 34, and confirmed using ground-based transit and radial velocity data. \textit{TESS} has now re-observed the system in Sector 61, and as shown in this work, the transits were observed several minutes earlier than expected if the period were constant.

In this paper we use these data, along with ground-based light curves and precise radial-velocity data, to explore the possibilities of orbital decay, apsidal precession, the R{\o}mer effect, and close companion planets as mechanisms to explain the observed timing variations. The rest of this paper is organized as follows. In Section \ref{sec:data} we analyze and process the transit light curves and radial velocity measurements. In Section \ref{sec:analysis} we fit the transit timing variations and radial velocity data to different models. In Section \ref{sec:conclusions} we present our conclusions.

\section{Data} \label{sec:data}
\subsection{TESS Photometry}

\textit{TESS} observed TOI-2818 in Sectors 7, 8, 34 and 61, providing a baseline that extends from 2019 January 7 to 12 February 2023. We used the shortest cadence data available for each sector. In years 1 and 3 of the mission, the target only appears in the \textit{TESS} full-frame images (FFIs). We used light curves at 1800-second cadence for Sectors 7 and 8 and 600-second cadence for Sector 34 from the \textit{TESS} Quick Look Pipeline (QLP) \citep{Huang_2020_QLPa, Huang_2020_QLPb, Kunimoto_2021_QLP}. In year 5 \textit{TESS} observed TOI-2818 at 120-second cadence; for Sector 61 we used the pre-search data conditioning simple aperture photometry light curves produced by the Science Processing Operations Center pipeline \citep{Smith_2012, Stumpe_2014, Jenkins_2016}. We downloaded both datasets from the Milkulski Archive for Space Telescopes \citep{https://doi.org/10.17909/t9-nmc8-f686, https://doi.org/10.17909/0cp4-2j79}. We selected a one-day long section around each transit to separate the transits and fit them individually.

\subsection{Ground-based Photometry}

\citet{Yee_2023} described observations of two transits of TOI-2818\,b by ground-based telescopes on 2021 December 8 and 2021 December 12 using the 0.51 m telescope at the El Sauce Observatory and a 0.4 m telescope as part of the Las Cumbres Observatory at the Cerro Tololo Inter-American Observatory (CTIO) respectively. These data are publicly available through ExoFOP \citep{exofop5}\footnote{\url{https://exofop.ipac.caltech.edu/tess}}.

\subsection{Radial Velocity Observations}
\citet{Yee_2023} report 7 radial velocity measurements of TOI-2818 recorded between 2021 December 21 and 2022 March 15 by the CTIO High Resolution Spectrometer (CHIRON; \citet{Tokovinin_2013, Paredes_2021}) on the CTIO 1.5 m telescope. We removed one observation from our analysis as it was collected during the transit of TOI-2818\,b, to avoid the impact of the Rossiter-McLaughlin effect on the measured radial velocity. We collected two new measurements on 2023 June 2 and 2023 June 4 following the same methodology.

We obtained eight radial velocity measurements taken with VLT/ESPRESSO (Echelle SPectrograph for Rocky Exoplanets and Stable Spectrographic Observations; \citet{Pepe_2021}). Seven observations were taken between 2023 June 19 and 2023 June 28, and an eighth on 2023 September 5, under program ID 111.264Q (PI Montet). All exposures were 600 seconds in length, using a sky fiber for background subtraction and read out in the slow readout mode with 2x1 binning. We infer the stellar RV and its uncertainty at each observation time through the ESPRESSO DRS pipeline v3.2.0. These radial velocity measurements are recorded in Table \ref{tab:radial_velocities}.

\begin{deluxetable}{lclr}

\tablecaption{Radial Velocity Measurements of TOI-2818}
\label{tab:radial_velocities}
\tablewidth{0pt}
\tablehead{
\colhead{Source} & \colhead{Time} & \colhead{RV} & \colhead{Uncertainty}\\
\colhead{} & \colhead{(BJD$_{\mathrm{TDB}}$)} & \colhead{(m s$^{-1}$)} & \colhead{(m s$^{-1}$)}
}
\decimals
\startdata
CHIRON$^1$   & 2459569.7244 & 58907$\phd\phn$ & 30$\phd\phn$ \\
CHIRON$^1$   & 2459571.7617 & 59038$\phd\phn$ & 30$\phd\phn$ \\
CHIRON$^1$   & 2459595.7596 & 59071$\phd\phn$ & 23$\phd\phn$ \\
CHIRON$^1$   & 2459597.7544 & 58868$\phd\phn$ & 27$\phd\phn$ \\
CHIRON$^1$   & 2459620.5707 & 59093$\phd\phn$ & 33$\phd\phn$ \\
CHIRON$^1$   & 2459622.5622 & 58920$\phd\phn$ & 27$\phd\phn$ \\
CHIRON$^{1,2}$& 2459653.5862 & 59081$\phd\phn$ & 24$\phd\phn$ \\
CHIRON       & 2460097.5217 & 59082$\phd\phn$ & 26$\phd\phn$ \\
CHIRON       & 2460099.5086 & 59023$\phd\phn$ & 51$\phd\phn$ \\
VLT/ESPRESSO & 2460115.4623 & 60380.8 &  1.8 \\
VLT/ESPRESSO & 2460116.4475 & 60473.9 &  2.6 \\
VLT/ESPRESSO & 2460117.4540 & 60519.4 &  2.8 \\
VLT/ESPRESSO & 2460118.4547 & 60407.9 &  8.3 \\
VLT/ESPRESSO & 2460122.4548 & 60417.9 &  6.2 \\
VLT/ESPRESSO & 2460123.4499 & 60368.8 &  2.3 \\
VLT/ESPRESSO & 2460124.4581 & 60463.6 &  2.1 \\
VLT/ESPRESSO & 2460192.8707 & 60438.7 &  5.2 \\
\tableline
$\gamma_{\mathrm{CHIRON}}$       &       & 58993$\phd\phn$ & 11$\phd\phn$ \\
$\gamma_{\mathrm{VLT/ESPRESSO}}$ &       & 60445.0 & $1.5$ \\
\enddata
\tablecomments{$^1$ Data from \citet{Yee_2023}. $^2$ Value measured during transit, removed from analysis due to Rossiter-Mclaughlin effect.}
\end{deluxetable}

\subsection{Model Fitting}

We fit the light curves following a similar methodology to that described in \citet{McKee_2023} using the python \texttt{exoplanet} package \citep{exoplanet:joss}. We fit each \textit{TESS} transit using common limb darkening coefficients $u_1$ and $u_2$ following the reparameterization of \citet{Kipping_2013}, while each ground-based transit used different limb darkening coefficients.

\begin{figure}[htb!]
\includegraphics[width=\columnwidth]{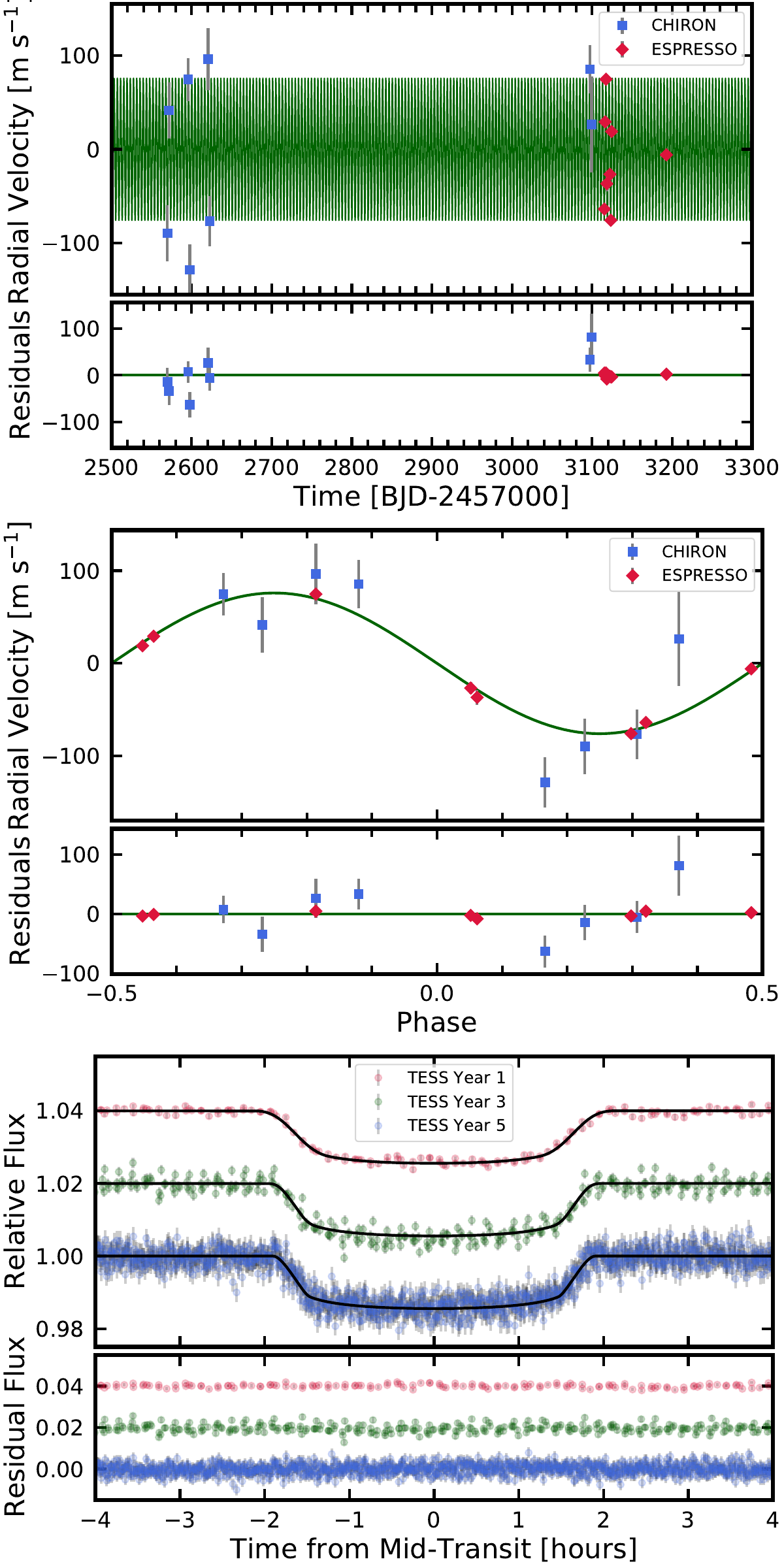}
\caption{The top panels show the radial velocities measured by CHIRON (blue) and VLT/ESPRESSO (red) compared to the best-fitting circular orbit model over the observation baseline, the panel below shows the residuals. The middle panel shows the radial velocities folded over the phase of the orbit, with the residuals below. A circular orbit is a good fit to the data. The bottom panels show the light curves of the transits aligned to the mid-transit times. Each year is stacked together, with vertical offsets between years 1 (red), 3 (green) and 5 (blue). The panel below shows the residuals to the light curve fit in the same fashion. The depth and duration of the transit does not change between each year of observations.}
\label{fig:rv_combined}
\end{figure}

We sampled the planet radius $R_b$ and impact parameter $b$ from log uniform and uniform priors respectively; these parameters are shared for each transit. We assigned priors for the mass $M_\star$ and radius $R_\star$ of the star as inferred from the spectral analysis of \citet{Yee_2023}. The mean out-of-transit flux value was allowed to vary for each transit. We included a quadratic model for the out of transit flux in the ground-based photometry to account for atmospheric effects. Given the proximity of the planet to the star, and the theoretical presumption that the eccentricity has been damped to an undetectably low level, we fix the eccentricity of the orbit to 0 for the light curve fitting. We allowed the mid-transit time of each transit to be a free parameter, in order to search for TTVs. We sampled each posterior distribution with \texttt{PyMC3} \citep{exoplanet:pymc3}, with 8,000 tuning steps and 40,000 draws. We verified our chains converged via visual inspection and following the diagnostic of \citet{Geweke92}. The mid-transit times found are listed in Table \ref{tab:transit_times}. The light curve fit to each year of transits is presented in Figure \ref{fig:rv_combined}.
\begin{deluxetable}{lccr}
\tablecaption{Transit Times for TOI-2818\,b}
\label{tab:transit_times}
\tablewidth{0pt}
\tablehead{
\colhead{Source} & \colhead{Transit Time} & \colhead{Uncertainty} & \colhead{Epoch}\\
\colhead{} & \colhead{(BJD$_{\mathrm{TDB}}$)} & \colhead{(days)} & \colhead{}
}
\startdata
\textit{TESS} Sector  7 & 2458494.2454 & 0.0011 &   0 \\
\textit{TESS} Sector  7 & 2458498.2820 & 0.0011 &   1 \\
\textit{TESS} Sector  7 & 2458502.3240 & 0.0011 &   2 \\
\textit{TESS} Sector  7 & 2458506.3639 & 0.0011 &   3 \\
\textit{TESS} Sector  7 & 2458510.4043 & 0.0012 &   4 \\
\textit{TESS} Sector  7 & 2458514.4414 & 0.0010 &   5 \\
\textit{TESS} Sector  8 & 2458518.4830 & 0.0011 &   6 \\
\textit{TESS} Sector  8 & 2458522.5208 & 0.0012 &   7 \\
\textit{TESS} Sector  8 & 2458530.6008 & 0.0012 &   9 \\
\textit{TESS} Sector  8 & 2458538.6828 & 0.0012 &  11 \\
\textit{TESS} Sector 34 & 2459229.4745 & 0.0009 & 182 \\
\textit{TESS} Sector 34 & 2459233.5141 & 0.0010 & 183 \\
\textit{TESS} Sector 34 & 2459237.5519 & 0.0009 & 184 \\
\textit{TESS} Sector 34 & 2459245.6328 & 0.0010 & 186 \\
\textit{TESS} Sector 34 & 2459249.6702 & 0.0009 & 187 \\
\textit{TESS} Sector 34 & 2459253.7110 & 0.0009 & 188 \\
El Sauce                & 2459556.6885 & 0.0006 & 263 \\
LCO CTIO                & 2459560.7284 & 0.0021 & 264 \\
\textit{TESS} Sector 61 & 2459964.6962 & 0.0008 & 364 \\
\textit{TESS} Sector 61 & 2459968.7376 & 0.0007 & 365 \\
\textit{TESS} Sector 61 & 2459972.7753 & 0.0008 & 366 \\
\textit{TESS} Sector 61 & 2459980.8553 & 0.0008 & 368 \\
\textit{TESS} Sector 61 & 2459984.8947 & 0.0008 & 369 \\
\enddata
\end{deluxetable}

We jointly fit a circular orbit to the TTVs and the RVs using the python \texttt{TTVFast} package \citep{Deck_2014}. We use \texttt{TTVFast} here to remain consistent with later models which model interactions between multiple planets. We assign a mass $M$, period $P$, longitude of the ascending node $\Omega$, and mean longitude $\lambda$ to the planet. We fix the longitude of the ascending node for the planet as $\Omega_b=0$, as it can not be inferred from the available data. We fix the inclination to $i = 88.3\degr$, calculated from the best fit light curve model. The stellar mass is fit with the same prior as previously, for all other parameters we apply a uniform prior. We fit independent offsets for the two sources of radial velocities. We also included jitter terms added in quadrature to the uncertainties from \textit{TESS} years 1, 3 and 5 separately, each ground-based transit, and CHIRON and ESPRESSO RVs. The model is started at a time of BJD 2458494. This model is the basis of the linear ephemeris used in Figure \ref{fig:ttv_combined}. The fit of the radial velocities is shown in Figure \ref{fig:rv_combined}.

We also fit a model with a linear acceleration term to the radial velocity to identify long-term trends, finding $\dot{v}_r = +0.08 \pm 0.05$ m s$^{-1}$ day$^{-1}$. Table \ref{tab:results} gives the mass and period of the planet found with and without an acceleration term.

\begin{deluxetable}{lrl}

\tablecaption{Parameters at BJD 2458494}
\label{tab:results}
\tablewidth{0pt}
\tablehead{
\colhead{Parameter} & \colhead{Value} & \colhead{Uncertainty}\\
}
\decimals
\startdata
\textbf{Constant Period Model} \\
Period (days) & 4.039705 & $^{+0.000006}_{-0.000002}$ \\
Planet Mass ($M_{\mathrm{Jup}}$) & 0.61 & $\pm0.03$ \\
\tableline
\textbf{Linear Acceleration Model} \\
Period (days) & 4.039706 & $^{+0.000007}_{-0.000002}$ \\
Planet Mass ($M_{\mathrm{Jup}}$) & 0.60 & $\pm0.03$ \\
Acceleration (m s$^{-1}$ day$^{-1}$) & 0.08 & $\pm0.05$\\
\tableline
\textbf{Decaying Period Model} \\
Period (days) & 4.039736 & $\pm0.000006$ \\
Planet Mass ($M_{\mathrm{Jup}}$) & 0.60 & $\pm0.03$ \\
Decay Rate (days orbit$^{-1}\times10^{-7}$) & $-1.7$ & $\pm0.3$\\
Tidal Quality Factor $Q'_*$ & $10.1$ & $^{+2.2}_{-1.5}$\\
R{\o}mer Acceleration (m s$^{-1}$ day$^{-1}$) & -3.2 & $\pm0.5$ \\
\tableline
\textbf{Outer Companion Model} \\
Period (days) & 4.039697 & $^{+0.000006}_{-0.000004}$ \\
Planet Mass ($M_{\mathrm{Jup}}$) & 0.61 & $^{+0.06}_{-0.03}$ \\
Eccentricity & 0.014 & $^{+0.005}_{-0.007}$ \\
Argument of Periastron (deg) & $-24$ & $\pm25$\\
Companion Period (days) & 2150 & $^{+330}_{-350}$ \\
Companion Mass ($M_{\mathrm{Jup}}$) & 71 & $^{+35}_{-30}$ \\
Companion Eccentricity & 0.84 & $^{+0.07}_{-0.12}$ \\
Companion Argument (deg) & $9$ & $^{+9}_{-8}$\\
\tableline
\textbf{Apsidal Precession Model} \\
Period (days) & 4.039709 & $^{+0.000022}_{-0.000010}$ \\
Planet Mass ($M_{\mathrm{Jup}}$) & 0.61 & $\pm0.03$ \\
Eccentricity & 0.004 & $^{+0.004}_{-0.003}$ \\
Argument of Periastron (deg) & $-50$ & $^{+40}_{-70}$\\
\tableline
\textbf{Best 16-Day Planet Model} \\
Period (days) & 4.03972459 \\
Planet Mass ($M_{\mathrm{Jup}}$) & 0.493477\\
Eccentricity & 0.0115 \\
Argument of Periastron (deg) & -38.18\\
Companion Period (days) & 16.1953744 \\
Companion Mass ($M_{\oplus}$) & 13.976203 \\
Companion Eccentricity & 0.0671 \\
Companion Argument (deg) & 47.87\\
\enddata
\end{deluxetable}


\section{Analysis} \label{sec:analysis}

\subsection{Tidal Inspiral}

A decaying orbit with a period that decreases at a constant rate will produce quadratic TTVs. We generated transit times and RVs for a circular orbit with a constant period with \texttt{TTVFast}, then applied a correction to the transit times to simulate a decaying period. The correction factor added is
\begin{equation}
\Delta t=\frac{1}{2}\frac{dP}{dN}N^2,
\label{eq:decay}
\end{equation}
where $P$ is the period of the planet and $N$ the transit epoch, following \citet{Patra_2017}.
We sampled with \texttt{emcee} \citep{Foreman_Mackey_2013}, an implementation of the affine invariant MCMC sampler of \citet{MCMC}, with the maximum-likelihood fit used as starting conditions for an N-body integration. We sample with 100 walkers and 10,000 steps, with the first 9,000 discarded as burn-in. The results of this fitting are listed in Table \ref{tab:results}.

\begin{figure*}[htb!]
\includegraphics[width=\linewidth]{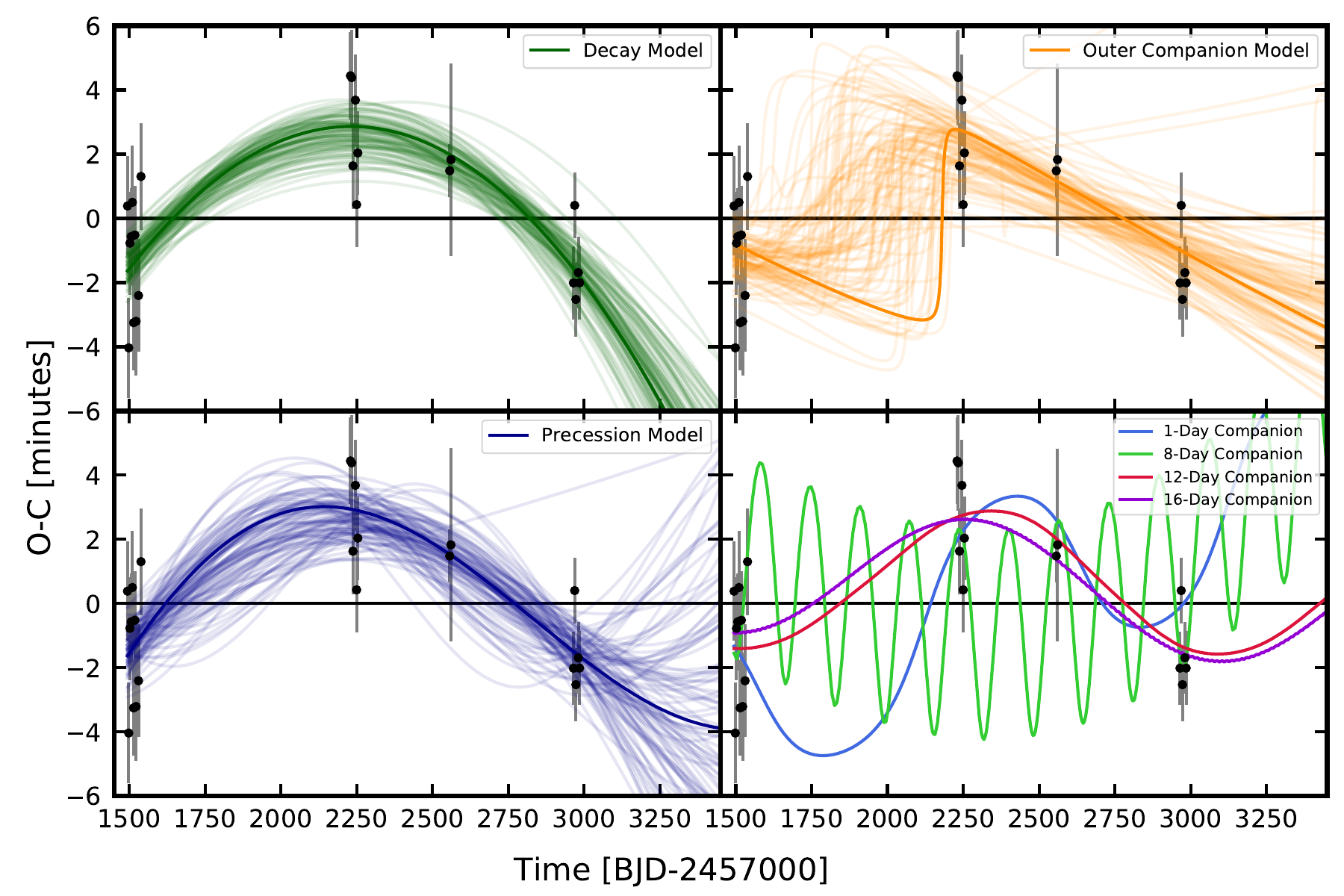}
\caption{TTV models are compared to the measured mid-transit times that are displayed as black points. The black line represents a linear ephemeris. The top left panel shows representative samples of the decay model in transparent green, along with the best fit as a solid line. The top right and bottom left panels show the same for the outer companion acceleration model (orange) and apsidal precession model (blue) respectively. The bottom right panel show the best fitting models featuring companion planets at 1 (blue), 8 (green), 12 (red) and 16 days (purple). Each type of model can fit the TTVs, but other constraints make all but the planetary model physically implausible.}
\label{fig:ttv_combined}
\end{figure*}

The TTVs produced are displayed in Figure \ref{fig:ttv_combined}. We find a best-fitting decay rate of $1.35^{+0.24}_{-0.26}$ s yr$^{-1}$. This is more than an order of magnitude larger than the decay rate for both WASP-12\,b \cite[$29\pm2$ ms yr$^{-1}$,][]{Yee_2020} and Kepler-1658\,b \cite[$131^{+20}_{-22}$ ms yr$^{-1}$,][] {Vissapragada_2022}.
The decay timescale of this system under this model, $P/\dot{P}$, is $270^{+70}_{-50}$ kyr, a factor of 10 shorter than the other planets.

If the cause of this decay is tidal dissipation then within the framework of the simplified tidal model of \citet{Goldreich_1966} we can determine the modified tidal quality factor, $Q'_*$, of the planet as
\begin{equation}
\dot{P}=-\frac{27\pi}{2Q'_*}\left(\frac{M_p}{M_*}\right)\left(\frac{R_*}{a}\right)^5,
\label{eq:quality_factor}
\end{equation}
where $a$ is the semi-major axis of the planet's orbit. For this system we calculate a tidal quality factor of
\begin{equation}
Q'_*=10.1^{+2.2}_{-1.5}.
\label{eq:q_value}
\end{equation}
This result is orders of magnitude smaller than the values reported for WASP-12\,b and Kepler-1658\,b of $1.75^{+0.13}_{-0.11}\times10^5$ and $2.50^{+0.85}_{-0.62}\times10^4$ respectively. The tidal quality is typically at least $10^5-10^7$ for hot Jupiters, but can be lower if the star is a subgiant such as WASP-12 \citep{Weinberg_2017}. TOI-2818 is closer to the main sequence than WASP-12. Nonlinear effects such as wave breaking can also result in smaller tidal quality factors, although not as low as 10 \citep{Barker_2020}. 

\subsection{R{\o}mer Delay}

The same quadratic TTV effect can be observed if the star is being accelerated steadily along the line of sight towards the observer due to the R{\o}mer effect. A constant acceleration causes an apparent change in period given by
\begin{equation}
\frac{\dot{P}}{P}\approx \frac{\dot{v}_r}{c},
\label{eq:romer_effect}
\end{equation}
where $\dot{v}_r$ is the change in line of sight velocity. The implied acceleration of TOI-2818 is $-3.2^{+0.5}_{-0.5}$ m s$^{-1}$ day$^{-1}$. This is significantly larger than the $\dot{v}_r = +0.08^{+0.05}_{-0.05}$ m s$^{-1}$ day$^{-1}$ inferred from Doppler spectroscopy as calculated in Section \ref{sec:data}.

\vspace{4em}

\subsection{Acceleration by an Outer Companion}

A constant acceleration throughout the timespan of the data has been ruled out by the lack of an RV trend. Very wide binaries beyond $\sim300$ AU are excluded by high-resolution imaging \citep{Yee_2023}. A nearer (few AU) companion to the system would produce an acceleration that is not necessarily constant over the timespan of the data; instead inducing Keplerian motion of the star/planet system. The acceleration will cause the observed transits to appear closer together or further apart as the light travel time changes at a non-constant rate. This more complex model might be compatible with the TTVs and radial velocity measurements, given the uneven time sampling of both types of data.

To test this we include a $\sim50M_\mathrm{Jup}$ brown dwarf in our \texttt{TTVFast} model, allowing the same parameters as for the planet to vary, with the addition of initial eccentricity $e$, and argument of periapsis $\omega$. In order to determine the TTVs, we calculate the light travel time delay between the star and the center of mass of the system. \texttt{TTVFast} does not account for this effect, and does not provide the three-dimensional coordinates of the star at the time of each transit. Instead we integrate the radial velocity of the star to infer the radial position of the star. We convert this distance into light travel time and correct the transit times. We again sample with \texttt{emcee} \citep{Foreman_Mackey_2013}, in the same manner as previously. The results of this fitting are listed in Table \ref{tab:results}. Samples of the model are shown in Figure \ref{fig:ttv_combined}.

We find that the results favor a companion of $\sim70M_\mathrm{Jup}$ and a $\sim2,100$-day period. The orbit has $e\approx0.8$, and an argument of periapsis nearly perpendicular to the line of sight at an angle of $9\pm9\degr$. This allows the radial velocities to have little variation across the observation interval, consistent with the lack of RV trend over our observing baseline. The companion induces an acceleration of $\sim60$ m s$^{-1}$ yr$^{-1}$, with a short RV shift up to $\sim2500$ m s$^{-1}$ at periastron. Additional radial velocity observations with CHIRON or ESPRESSO in particular could quickly detect or rule out such a companion.

This orbital configuration avoids RV detection by shifting motion from the line of sight to the plane of the sky. This should produce an astrometric signal that could be measured by \textit{Gaia} \citep{Gaia_2016}. The renormalized unit weight error (RUWE) is a measure of how well a single-star model fits to the data measured by \textit{Gaia} \citep{Lindegren_2018, Lindegren_2021}. This measure is sensitive to unresolved stellar companions \citep{Belokurov_2020}. The RUWE is expected to be around 1.0 for a good fit, a higher value may indicate that the source is not a single star. The RUWE from \textit{Gaia} DR3 for this system is 0.924 \citep{Gaia_DR3_2023} We simulated the expected RUWE for this system using the procedure described in \citet{Wallace_2024}. Simulating a brown dwarf in the requisite orbit produces a RUWE 0.924 or smaller only 7\% of the time, compared to 22\% of the time for single star models. The outer companion solution is therefore less likely to explain the RUWE than solutions that do not include a brown dwarf. The predicted RUWE is dependent on the time of periapse of the companion: due to the high eccentricity, there is relatively little orbital motion for much of the orbital period. By chance, Gaia DR3 observations do not contain the expected time of periapse; if the brown dwarf exists we expect the RUWE over a full orbit to be as high as 1.7. \textit{Gaia} DR4 should provide the astrometric data over a long enough baseline to detect or rule out this hypothetical companion.

Despite fitting the transits well, the high eccentricity and short orbit result in a close periastron of $<0.7$ AU. A large companion this close to the host star will truncate the protoplanetary disk, presenting challenges for Jupiter-sized planet formation \citep{Jang_Condell_2015}. It is possible that the current orbital configuration was not present at formation and that the eccentricity of the brown dwarf was excited over time, possibly through dynamical interactions with another companion. The close passes of the companion can introduce dynamic instability to the system, which may not survive. The best-fitting solution is stable over a timescale of 100,000 years, however the Lyapunov time is only 200 years, suggesting that the system is chaotic.

In summary, an outer companion can explain the observations, but requires specific orbital configurations and coincidences in observation timing to evade detection. Additional RV or astrometric data should be conclusive about the possibility of a brown dwarf in this system as the cause of the TTVs.

\subsection{Apsidal Precession}

Apsidal precession can create sinusoidal TTVs due to the change of orientation of the elliptical orbit of the planet in time. We use the same method as previously described by using \texttt{TTVFast} to create an eccentric orbit with constant period, then applying a correction factor to the transit times. The correction applied is:
\begin{equation}
\Delta t=-\frac{eP_a}{\pi}\cos{\omega(N)},
\label{eq:apsidal_precession}
\end{equation}

\begin{equation}
\omega(N)=\omega_0+\frac{d\omega}{dN}N,
\label{eq:precession_term}
\end{equation}

\begin{equation}
P_s=P_a\left(1-\frac{1}{2\pi}\frac{d\omega}{dN}\right),
\label{eq:period_term}
\end{equation}

as described by \citet{Gimenez_1995}, and adapted to the low eccentricity, high inclination limit by \citet{Patra_2017}. We again sample with \texttt{emcee} \citep{Foreman_Mackey_2013},  in the same manner as previously. The results of this fitting are listed in Table \ref{tab:results}. We find that the best-fitting eccentricity is $0.004^{+0.004}_{-0.003}$ with a best-fitting precession rate of $38^{+25}_{-14}\degr$ yr$^{-1}$. The model is shown in Figure \ref{fig:ttv_combined}.

This precession rate is high and difficult to explain from a theoretical perspective. The theoretical precession rate in most hot Jupiter systems is thought to be dominated by the contribution from the planetary tidal bulge, which can be used to calculate the planet's Love number $k_{2p}$ \citep{Ragozzine_2009,Vissapragada_2022}:
\begin{equation}
\frac{d\omega}{dN}=15\pi k_{2p}\left(\frac{M_*}{M_p}\right)\left(\frac{R_p}{a}\right)^5.
\label{eq:love_number_planet}
\end{equation}

For a precession rate of $38^{+25}_{-14}\degr$ yr$^{-1}$ to be caused by planetary tidal bulges, the value of $k_{2p}$ is $270^{+180}_{-100}$. The Love number $k_2$ is a measure of the mass distribution inside a planet. The value is required to be between 0 and 1.5, with a homogeneous body yielding the maximum \citep{Kramm_2011}. This calculated value is much larger than the maximum, and so is unphysical.

If the star is evolved the stellar tidal bulge may be a larger contributor to precession. If we calculate the stellar Love number $k_{2*}$ using:
\begin{equation}
\frac{d\omega}{dN}=15\pi k_{2*}\left(\frac{M_p}{M_*}\right)\left(\frac{R_*}{a}\right)^5,
\label{eq:love_number_star}
\end{equation}

we find a value of $1.6^{+1.0}_{-0.6}\times10^5$, which is even larger. A higher eccentricity orbit decreases the required precession rate, so we can check whether any physical solution for $k_{2p}$ can exist as $e$ approaches 1. Using Equation \ref{eq:love_number_planet}, we can find that the precession rate is $k_{2p}\times0.14\degr$ yr$^{-1}$. 
To create the largest TTVs, the cosine term of Equation \ref{eq:apsidal_precession} must have a large derivative. Here the small-angle approximation for the change in value of the cosine term is parabolic, equal to $\frac{1}{2}(\omega (N))^2$. 
Across half the TTV baseline, the most that the planet can precess is $k_{2p}\times0.28\degr$, or $k_{2p}\times0.005$ radians, implying the cosine term can change by at most $(k_{2p})^2\times1.25\times10^{-5}$. 
Using Equation \ref{eq:apsidal_precession}, we find $\Delta t = e\times(k_{2p})^2\times1.6\times10^{-5}$ days, or $e\times(k_{2p})^2\times0.023$ minutes. 
To get $\Delta t$ of $\sim4$ minutes, $e\times(k_{2p})^2\sim170$.
As the Love number has an upper bound of 1.5, this is not possible for any eccentricity.

\citet{Yee_2023} report a circularization timescale of $\sim0.2$ Gyr and a stellar age of $\sim9.5$ Gyr. Using Equation \ref{eq:apsidal_precession}, an eccentricity of $\sim0.001$ will result in a TTV amplitude of $\sim4$ minutes, so the eccentricity must be at least this value to produce the observed TTVs, regardless of precession rate. A highly eccentric orbit should be circularized below this value over a time period of 1-2 billion years. Without a mechanism to increase it, the eccentricity will be too small for apsidal precession to account for the observed TTVs.

Precession can also be caused by stellar oblateness \citep{Sterne_1939}. For a circular, aligned orbit the precession can be calculated using:
\begin{equation}
\frac{d\omega}{dN}=\frac{3}{2} J_{2}\left(\frac{R_*}{a}\right)^2,
\label{eq:stellar_oblateness}
\end{equation}
where $J_2$ is the gravitational potential due to oblateness \citep{Li_2020}. This value is typically $\sim10^{-7}$. Here we find an asymmetric distribution for $J_2$ of $\log_{10}(J_2)=-0.42^{+0.24}_{-0.19}$, many orders of magnitude larger, suggesting that oblateness cannot cause the required precession.

\vspace{1em}

\subsection{A Near-Resonant Planet}

Another possible cause of the TTVs is perturbations from a low-mass, nearby planet. Such a planet needs a sufficiently low mass to evade detection in the ESPRESSO data. A planet with a small radius and large TTVs may be difficult to detect in \textit{TESS} data even if transiting. In order to test this possibility we simulate a planet with an orbital period near different inner and outer mean-motion resonances of TOI-2818\,b. These configurations induce detectable TTVs even for a small planet, and cause the TTVs to have a longer super-period in their trends in order to cover the long \textit{TESS} baseline.

The resulting fits are poorly constrained by the lack of information about the second planet. Without further transits or additional RV measurements, a unique solution cannot be obtained. We present multiple best fit solutions for different orbital periods to demonstrate the plausibility of a planet as the cause of the TTVs. In all tested periods, a planet in the $\lesssim 10 M_\oplus$ mass range is sufficient to reproduce the TTVs; these planets would not be seen in ESPRESSO data. An example given by the best-fitting 16-day planet model is listed in Table \ref{tab:results}. This features the largest planet at $\sim 10 M_\oplus$, while the similar TTV curve of the 12-day solution has a $< 1 M_\oplus$ planet. The 16-day solution decreases the mass of planet b by $0.1 M_{\mathrm{Jup}}$, while the other solutions do not significantly change the mass of the hot Jupiter. The best fitting TTV model for planets on 16-, 12-, 8- and 1-day orbits are shown in Figure \ref{fig:ttv_combined}.

We find that as the orbit of the outer planet becomes wider, the change in TTVs becomes slower. The 12- and 16-day orbits produce lower frequency TTVs that better match the \textit{TESS} observing cadence.

We integrate each of the best-fitting solutions to test the stability of these configurations with \texttt{REBOUND} \citep{rebound, reboundias15}. We find that these systems are stable over a period of at least 100,000 years.

\subsection{Constraints on Potential Moons}

The existence of a moon can create TTVs by moving the planet around the barycenter of the planet-moon system. To maximize the size of the TTVs, we consider a moon at half the Hill radius \citep{Domingos_2006}. In this system, half the Hill radius is $\approx$220,000 km. The orbital velocity of the planet is $\approx$8100 km min$^{-1}$. A moon of $\approx15M_\oplus$ at half the Hill radius is massive enough to shift the planet from the barycenter by 16,000 km, causing a TTV semi-amplitude of 2 minutes as observed. Such a small moon might not create an obvious transit signal on either side of the planet transit. For a sector of \textit{TESS} transits to be early, the moon must be in the same location for each transit. The orbital periods of the moon and planet must be very near an orbital commensurability, an unlikely but not impossible coincidence. Continued TTV monitoring could enable inference of a unique candidate moon solution which could be tested with high-precision photometric observations by \textit{HST} or \textit{JWST}.


\section{Conclusions} \label{sec:conclusions}

TOI-2818\,b shows TTVs that deviate from a constant period over a 4 year baseline of \textit{TESS} data. Multiple models are able to fit the transit times, but are inconsistent with either the radial velocity measurements, or our knowledge of planet structure and tides. The preferred model we investigated that can explain the data invokes gravitational perturbations from a small nearby planet, although the data are insufficient to find a single solution. More observations in the form of additional transits and high precision radial velocity measurements are necessary to confirm and characterize the second planet. 
The old age of this system is significant as it shows that such a system can exist after billions of year, without the orbits decaying or becoming unstable. This is suggestive of smooth disk migration leading to the current orbital configuration. The rates of hot Jupiters with and without TTVs due to companion planets will reveal the relative rates of formation through smooth disk migration, which will preserve companions in orbital resonances, and high eccentricity migration, which will lead to the scattering of companions. This system, a hot Jupiter with a likely companion around an evolved star with a precisely known age, can provide a benchmark to better understand the lifetimes of hot Jupiters and their environments.

\section*{Acknowledgments}

We thank João Faria for constructive discussions that improved the quality of this manuscript.

We thank the anonymous referee for their prompt and helpful feedback.

This research made use of \texttt{exoplanet} \citep{exoplanet:joss,
exoplanet:zenodo} and its dependencies \citep{exoplanet:agol20,
exoplanet:arviz, exoplanet:astropy13, exoplanet:astropy18, exoplanet:luger18,
exoplanet:pymc3, exoplanet:theano}.

This paper includes data collected with the TESS mission, obtained from the MAST data archive at the Space Telescope Science Institute (STScI). The specific observations analyzed can be accessed via \dataset[https://doi.org/10.17909/0cp4-2j79]{https://doi.org/10.17909/0cp4-2j79} \citep{https://doi.org/10.17909/0cp4-2j79} and \dataset[https://doi.org/10.17909/t9-nmc8-f686]{https://doi.org/10.17909/t9-nmc8-f686} \citep{https://doi.org/10.17909/t9-nmc8-f686}. Funding for the TESS mission is provided by the NASA Explorer Program. STScI is operated by the Association of Universities for Research in Astronomy, Inc., under NASA contract NAS 5–26555.

This research has made use of the Exoplanet Follow-up Observation Program website, which is operated by the California Institute of Technology, under contract with the National Aeronautics and Space Administration under the Exoplanet Exploration Program.

This research has used data from the SMARTS 1.5m telescope, which is operated as part of the SMARTS Consortium. Based in part on observations at NSF Cerro Tololo Inter-American Observatory, NSF NOIRLab, which is managed by the Association of Universities for Research in Astronomy (AURA) under a cooperative agreement with the U.S. National Science Foundation.

Based on observations collected at the European Southern Observatory under ESO programme 111.264Q (PI Montet).

This work has made use of data from the European Space Agency (ESA) mission {\it Gaia} (\url{https://www.cosmos.esa.int/gaia}), processed by the {\it Gaia} Data Processing and Analysis Consortium (DPAC, \url{https://www.cosmos.esa.int/web/gaia/dpac/consortium}). Funding for the DPAC has been provided by national institutions, in particular the institutions participating in the {\it Gaia} Multilateral Agreement.

B.J.M. is supported by an Australian Government Research Training Program (RTP) Scholarship.

J.H.C.M. is supported by FCT - Fundação para a Ciência e a Tecnologia through national funds by these grants: UIDB/04434/2020, UIDP/04434/2020, PTDC/FIS-AST/4862/2020. J.H.C.M. is also supported by the European Union (ERC, FIERCE, 101052347). 

A.M.S acknowledges support from the Fundação para a Ciência e a Tecnologia (FCT) through the Fellowship 2020.05387.BD (DOI: 10.54499/2020.05387.BD). Funded/Co-funded by the European Union (ERC, FIERCE, 101052347). Views and opinions expressed are however those of the author(s) only and do not necessarily reflect those of the European Union or the European Research Council. Neither the European Union nor the granting authority can be held responsible for them. This work was supported by FCT - Fundação para a Ciência e a Tecnologia through national funds by these grants: UIDB/04434/2020, UIDP/04434/2020.


\facilities{TESS,  El Sauce: 0.6m, LCOGT, CTIO:1.5m (CHIRON), VLT:Antu (ESPRESSO)}

\software{
arviz \citep{exoplanet:arviz},
Astropy \citep{exoplanet:astropy13, exoplanet:astropy18},
astroquery \citep{astroquery},
emcee \citep{Foreman_Mackey_2013},
exoplanet \citep{exoplanet:joss, exoplanet:zenodo},
Lightkurve \citep{lightkurve},
Matplotlib \citep{matplotlib},
NumPy \citep{numpy},
PyMC3 \citep{exoplanet:pymc3},
REBOUND \citep{rebound, reboundias15},
SciPy \citep{scipy},
starry \citep{exoplanet:luger18}
Theano \citep{exoplanet:theano},
TTVFast \citep{Deck_2014},
}

\bibliography{main}{}
\bibliographystyle{aasjournal}



\end{document}